\documentclass[epj,nopacs]{svjour}
\usepackage{multirow}
\usepackage{graphicx}
\usepackage{amssymb,amsbsy,amsmath,amsfonts}
\usepackage{slashed}
\usepackage[latin1]{inputenc}

\begin{document}
\title{$N^*$ states with hidden charm and a three-body nature}
\author{Brenda B. Malabarba\inst{1}\and K. P. Khemchandani\inst{2}\and A. Mart\'inez Torres\inst{1}}
\institute{Universidade de Sao Paulo, Instituto de Fisica, C.P. 05389-970, Sao Paulo, Brazil, \email{amartine@if.usp.br}\and Universidade Federal de Sao Paulo, C.P. 01302-907, Sao Paulo, Brazil}
\date{\today}
\abstract{
In this work we study the formation of $N^*$'s  as a consequence of the dynamics involved in the $ND\bar D^*-N\bar D D^*$ system when the $D\bar D^*-\bar D D^*$ subsystem generates $X(3872)$ in isospin 0 and $Z_c(3900)$ in isospin 1.  States with isospin $I=1/2$ and mass in the energy region $4400-4600$ MeV are found to arise with spin-parity $J^P=1/2^+$ and $3/2^+$, leading to predictions in this way of the existence of $N^*$ resonances with hidden charm and a three-body nature. We also discuss the possibility of the existence of $\Delta_c$ states, i.e., $\Delta$'s with hidden charm.
}



\maketitle

\section{Introduction}
Recently, the LHCb collaboration announced the existence of possible hidden charm pentaquarks in the $J/\psi p$ invariant mass distribution of the process $\Lambda^0_b\to J/\psi p K^-$~\cite{Aaij:2015tga,Aaij:2016phn,Aaij:2019vzc}. The observation of such states, denoted as $P^+_c(4312)$, $P^+_c(4440)$ and $P^+_c(4457)$, constitutes a turning point in the experimental search for signals related to exotic baryons, which had gradually reduced after the lack of evidence for the existence of $\Theta^+(1540)$ was reported in higher statistics experiments~\cite{Battaglieri:2005er,Miwa:2006if,Moritsu:2014bht,Shen:2016csu}~\footnote{See also Refs.~\cite{Torres:2010jh,MartinezTorres:2010zzb} for a possible theoretical explanation of the signal observed by the LEPS collaboration for $\Theta^+(1540)$.}. Indeed, continuing with the hunting of exotic baryons, more recently, the LHCb collaboration has claimed the existence of yet another pentaquark with hidden charm, similar to the above mentioned $P_c$ states, but with nonzero strangeness, whose mass is around 4459 MeV~\cite{wang}. Coming back to the discussion on the $P_c$'s, we must recall that two states were initially claimed in Ref.~\cite{Aaij:2015tga}, the so called $P^+_c(4380)$, with a width of $205\pm18\pm86$ MeV, and $P^+_c(4450)$, with a width of $39\pm5\pm19$. Out of the two states, the existence of the former was based more on a requirement for a good description of the data than on a direct observation~\cite{Aaij:2015tga}. The best fit to the data in Ref.~\cite{Aaij:2015tga} was obtained by considering spin-parity $J^P=3/2^-$ for $P^+_c(4380)$ and $5/2^+$ for $P^+_c(4450)$, although acceptable solutions were also obtained for the combinations $J^P=3/2^+$ and $5/2^-$ and $J^P=5/2^+$ and $3/2^-$, respectively. The updated analysis of the same processes as in Ref.~\cite{Aaij:2015tga}, but with much larger statistical significance, was made in Ref.~\cite{Aaij:2019vzc}, which revealed the presence of two states, $P^+_c(4440)$ (mass $M=4440.3\pm1.3^{+4.1}_{-4.7}$ MeV, width $\Gamma=20.6\pm 4.9^{+8.7}_{-10.1}$ MeV) and $P^+_c(4457)$ ($M=4457.3\pm0.6^{+4.1}_{-1.7}$ MeV, $\Gamma=6.4\pm 2.0^{+5.7}_{-1.9}$ MeV), instead of $P^+_c(4450)$, and a narrow peak at 4312 MeV, $P^+_c(4312)$ ($M=4311.9\pm0.7^{+6.8}_{-0.6}$ MeV, $\Gamma=9.8\pm 2.7^{+3.7}_{-4.5}$), though no definite spin-parity is assigned to the three states yet. The fits performed in Ref.~\cite{Aaij:2019vzc} with and without $P^+_c(4380)$ describe well the data, and its existence could not be ascertained. Several theoretical descriptions, as well as different spin-parity assignments, have been proposed for the nature of these $P_c$ states, like pentaquarks~\cite{Maiani:2015vwa,Lebed:2015tna,Ghosh:2015ksa,Weng:2019ynv,Wang:2019got}, meson-baryon states~\cite{Roca:2015dva,Xiao:2019aya,Liu:2019tjn,Burns:2019iih,He:2019ify,Xiao:2019mst,Guo:2019kdc,Du:2019pij,Xu:2020gjl}, peaks arising from triangle singularities~\cite{Guo:2015umn,Liu:2015fea,Nakamura:2021qvy} (which seems to be more plausible for $P^+_c(4457)$, although further investigations are necessary~\cite{Aaij:2019vzc}) or a virtual state for $P^+_c(4312)$~\cite{Fernandez-Ramirez:2019koa}. In case of the meson-baryon molecular attribution, the predominant  interpretation is to consider the $P_c$'s as states arising from the $\Sigma_c \bar D^*$, $\Sigma^*_c\bar D^*$ and coupled channel dynamics in s-wave. Within such a description, the $P_c$ states, although there is no consensus on the spin assignment, would have negative parity. However, in Ref.~\cite{Burns:2019iih}, by introducing $\Lambda_c(2595)\bar D$ coupled to $\Sigma_c\bar D^*$, the authors determined a positive parity assignment for $P_c(4457)$.

With all the experimental and theoretical efforts being made to understand the properties of the $P_c$ states claimed in Ref.~\cite{Aaij:2019vzc}, and with the debate on their spin-parity assignments still under way, it is important to investigate the possible formation of $N^*$ states with hidden charm and positive parity. Such states may naturally arise as a consequence of the dynamics involved in the three-body systems $ND\bar D^*$ and $N\bar D D^*$ since the interactions between the different subsystems are attractive in s-wave, forming states like $X(3872)$, $Z_c(3900)$, $\Lambda_c(2595)$~\cite{Braaten:2003he,AlFiky:2005jd,Gamermann:2007fi,Gamermann:2010zz,Aceti:2014uea,Ortega:2018cnm,He:2017lhy,Hofmann:2005sw,Mizutani:2006vq,GarciaRecio:2008dp,Romanets:2012hm,Liang:2014kra,Nieves:2019nol}. Further, the $ND\bar D^*$ ($N\bar D D^*$) threshold lies around 4814 MeV, and due to the strong attraction present in the $ND$ and $ND^*$ coupled channel system in s-wave, where $\Lambda_c(2595)$ (among other $\Lambda_c$ and $\Sigma_c$ states) has been found to get generated around 210 (352) MeV below the $DN$ ($D^*N$) threshold, we can expect large binding energies in the three-body system considered, of the order of 200-300 MeV. In this way, there exists a possibility of finding positive parity state(s) in the energy region 4400-4600 MeV, precisely where the $P_c$'s have been observed in Refs.~\cite{Aaij:2015tga,Aaij:2019vzc}. 

Motivated by this reasoning, in this work, we study the $NX(3872)-NZ_c(3900)$ coupled configurations of the $ND\bar D^*-N\bar D D^*$ system. To do this, we solve the Faddeev equations within the fixed center approximation (FCA)~\cite{Foldy:1945zz,Brueckner:1953zz,Deloff:1999gc,MartinezTorres:2020hus} assuming that $D\bar D^*-\bar D D^*$ clusters as $X(3872)/Z_c(3900)$. As we will show,  $N^*$ states with hidden charm, spin-parity $J^P=1/2^+$, $3/2^+$ and masses in the energy region 4400-4600 MeV are found to arise due to the underlying three-body dynamics. We discuss different processes where such positive parity $P_c$'s can be studied experimentally. We also investigate the possible  existence of isospin 3/2 states with hidden charm. Finally we test the applicability of  FCA to the current system by calculating diagrams beyond the approximation and show that such contributions are negligible.

\section{Formalism}\label{Fo}
\begin{figure}[h!]
\centering
\includegraphics[width=0.4\textwidth]{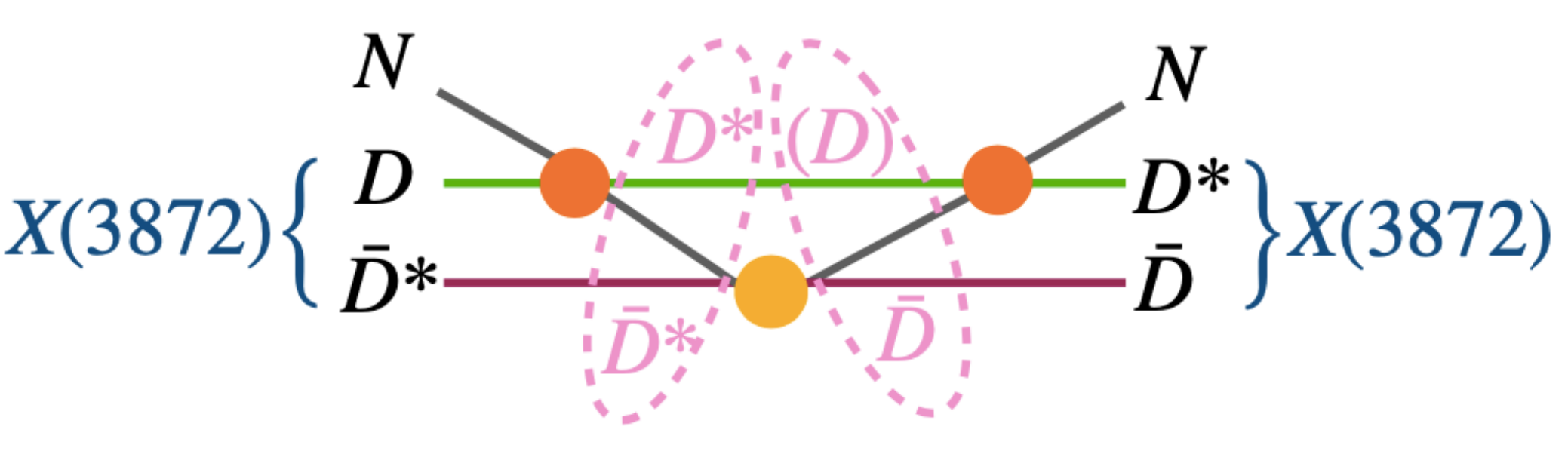}
\caption{A possible contribution to the transition $N(D\bar D^*)\to N(\bar D D^*)$ when $D\bar D^*$ ($\bar D D^*$) clusters as $X(3872)$. Reading the diagram from left to right, the nucleon interacts with the $D$ meson belonging to $X(3872)$ to produce $ND^*$ (or $ND$). The rescattered nucleon propagates and interacts with $\bar D^*$. After the latter interaction, the nucleon rescatters again and interacts with the $D^*$, or the $D$, meson formed in the first rescattering, leading to the final state $N(\bar D D^*)$. Such transition involves the presence of a virtual $N(D^*\bar D^*)$ or $N(D\bar D)$ states, which basically do not overlap with $NX(3872)$.}\label{break}
\end{figure}

In the FCA to the Faddeev equations, the scattering between the three particles forming the system is described in terms of a particle interacting with a scattering center. Such a treatment is relevant when two of the particles cluster as a state whose mass is heavier than the third one, which rescatters  with those forming the cluster through a series of interactions (for a review on the topic see Ref.~\cite{MartinezTorres:2020hus}). In this way, the cluster plays the role of the scattering center, which stays unaltered in the scattering process. In this work we study the $NX(3872)/NZ_c(3900)$ configurations of the $ND\bar D^*-N\bar D D^*$ system. We consider then that the $D\bar D^*-\bar DD^*$ system clusters as $X(3872)$ [$Z_c(3900)$] in isospin $0$ ($1$) and the nucleon rescatters, successively, off the $D$ ($\bar D$) and $\bar D^*$ ($D^*$) mesons. Since, 
\begin{align}
|X(3872)\rangle&=\frac{1}{\sqrt{2}}\Big[|D\bar D^*;I_{D\bar D^*}=0,I_{zD\bar D^*}=0\rangle\nonumber\\
&\quad+|\bar D D^*;I_{\bar DD^*}=0,I_{z\bar D D^*}=0\rangle\Big],\nonumber\\
|Z_c(3900)\rangle&=\frac{1}{\sqrt{2}}\Big[|D\bar D^*;I_{D\bar D^*}=1,I_{zD\bar D^*}=1\rangle\nonumber\\
&\quad+|\bar D D^*;I_{\bar DD^*}=1,I_{z\bar D D^*}=1\rangle\big],\nonumber
\end{align}
when adding a nucleon, we can write the three-body state as, 
\begin{align}
|N\mathbb{C}_a;I,I_z\rangle&=\frac{1}{\sqrt{2}}\Big[|N(D\bar D^*)_{I_a};I,I_z\rangle\nonumber\\
&\quad+|N(\bar D D^*)_{I_a};I,I_z\rangle\Big],
\end{align}
where $\mathbb{C}_a$ represents the cluster, whose isospin is $I_a$, and $I$ ($I_z$) is the isospin (and its third component) of the three-body system.
In this way, to describe the interaction between a $N$ and the cluster $\mathbb{C}_a$ we need to calculate the scattering $T$-matrix
\begin{align}
&\langle N\mathbb{C}_a;I,I_z|T|N\mathbb{C}_a;I,I_z\rangle\nonumber\\
&\quad=\frac{1}{2}\Big[\langle N(D\bar D^*)_{I_a};I,I_z|T|N(D\bar D^*)_{I_a};I,I_z\rangle\nonumber\\
&\quad\quad+\langle N(\bar D D^*)_{I_a};I,I_z|T|N(\bar D D^*)_{I_a};I,I_z\rangle\Big].\label{NCa}
\end{align}
Note that the transition
\begin{align}
\langle N(D\bar D^*)_{I_a};I,I_z|T|N(\bar D D^*)_{I_a};I,I_z\rangle \nonumber
\end{align}
requires that the scattering of the nucleon on $D\bar D^*$ converts the cluster to $D^*\bar D^*$ or $D\bar D$ in the intermediate states (see Fig.~\ref{break}). 
Since our purpose is to identify the cluster with $X(3872)$ or $Z_c(3900)$ in all intermediate states, both of which are understood as $D\bar D^*-\bar D D^*$ ``molecules'',   other intermediate processes  are expected to be small and are not considered in the formalism.
 Indeed it was shown in Refs.~\cite{MartinezTorres:2010ax,MartinezTorres:2020hus} that such considerations lead to small contributions to the dominant process for center of mass energies below the threshold. We thus limit ourselves to investigating the formation of three-body states below the threshold. Besides, to test the validity of the formalism, we have made explicit evaluation of the diagrams beyond FCA and confirm that such contributions are indeed negligible for the system considered in this work, at least for energies below the threshold (consistently with the findings of Ref.~\cite{MartinezTorres:2010ax}). Details of these calculations are given in a subsequent section.
\begin{figure}[h!]
\centering
\includegraphics[width=0.4\textwidth]{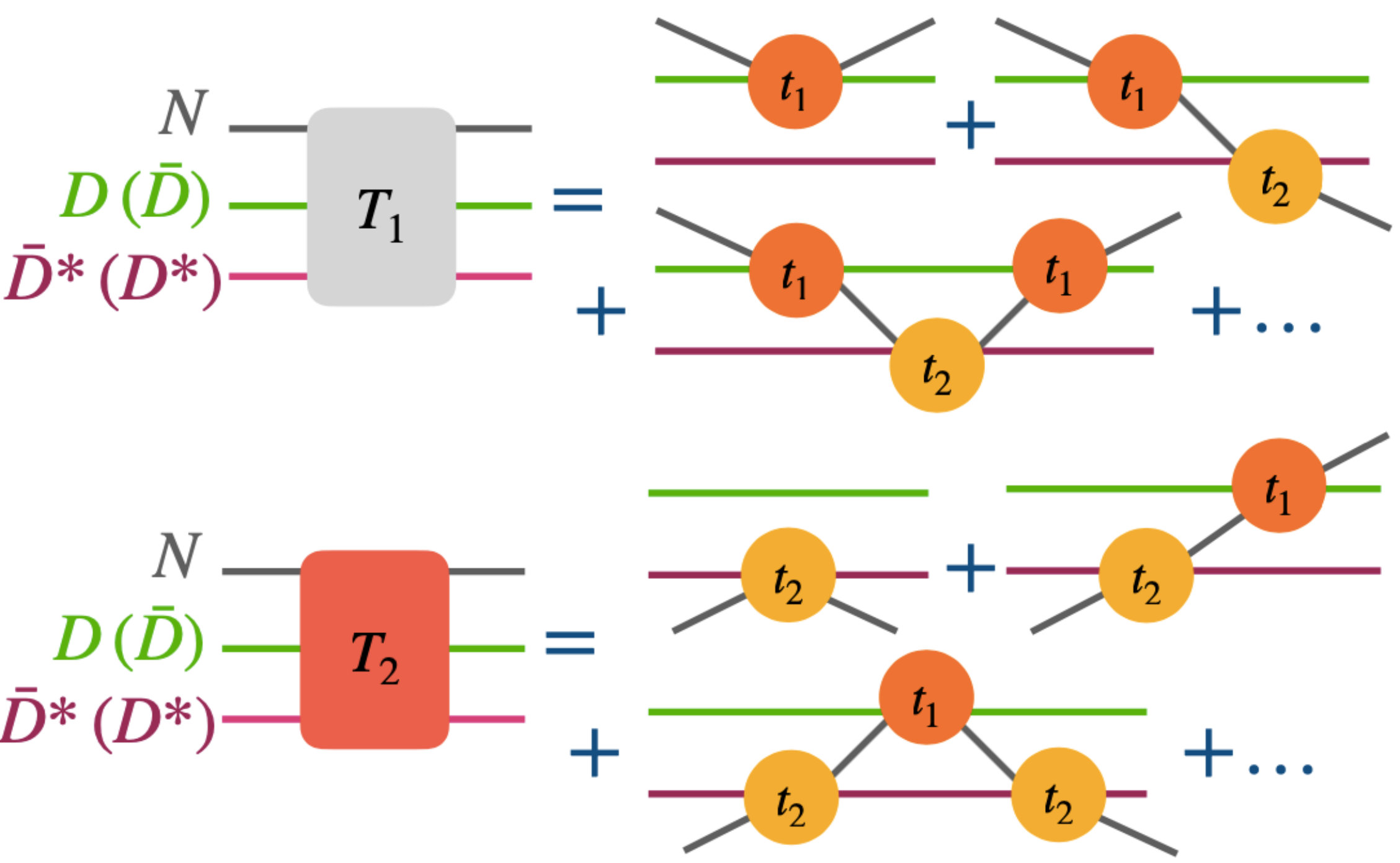}
\caption{Diagrammatic representation of some of the contributions to the $T_1$ and $T_2$ partitions.}\label{FCAseries}
\end{figure}

Continuing here with the discussions on the formalism, we can say that we need to study the $ND\bar D^*$ and $N\bar D D^*$ systems where $D\bar D^*$ ($\bar D D^*$) clusters either as $X(3872)$ or $Z_c(3900)$. Note that the interactions in the two kind of three-body systems are different. While one case involves a series of $ND$ and $N\bar D^*$ interactions, the other consists of $N\bar D$ and $N D^*$ interactions. Within the FCA to the Faddeev equations, the $T$-matrix for the $ND\bar D^*$ ($N\bar D D^*$) system is obtained from the resummation of two series which consider the rescattering of  the nucleon on $D$ ($\bar D$) and $\bar D^*$ ($D^*$), as shown in Fig.~\ref{FCAseries}. In this way, 
\begin{align}
T=T_1+T_2,\label{T}
\end{align}
where the partitions $T_1$ and $T_2$ are obtained by solving the coupled equations
\begin{align}
T_1&=t_1+t_1G_0T_2,\nonumber\\
T_2&=t_2+t_2G_0T_1,\label{Tpart}
\end{align}
for a given isospin $I$ and angular momentum $J$ of the three-body system. Since we consider s-wave interactions, the value of $J$ for the $ND\bar D^*$ ($N\bar D D^*$) system is given by the spin of the $N\bar D^*$ ($ND^*$) subsystem. The $t_1$ and $t_2$ in Eq.~(\ref{Tpart}) are two-body $t$-matrices related to the $ND$ ($N\bar D$) and $N\bar D^*$ ($ND^*$) systems, respectively, and $G_0$ represents the propagator of a nucleon in the cluster formed. For total isospin $I=1/2$ of the $ND\bar D^*$ ($N\bar D D^*$) system, we treat $N(D\bar D^*)_0$ and $N(D\bar D^*)_1$ [$N(\bar D D^*)_0$ and $N(\bar D D^*)_1$] as coupled channels, since both configurations can lead to isospin 1/2. 

We must emphasize here that both $X(3872)$ and $Z(3900)$ are interpreted as $D\bar D^*-\bar DD^*$ quasi-bound states with the same mass, although a different isospin and width~\cite{Gamermann:2007fi,Aceti:2014uea,Gamermann:2009uq}. Thus, FCA can be used to take into account transitions between the two systems as in the study of $Nf_0(980)-Na_0(980)$~\cite{Xie:2010ig}. In case a reader is interested in an analysis of the effects of  considering FCA for coupled channels with the clusters having  different masses, we refer the reader to Ref.~\cite{Roca:2011br}, where effective systems  like, $\pi K_2^*(1430)$ and $K a_2(1320)$ have been studied. 

In this way, $t_1$, $t_2$ and $G_0$ are matrices in the coupled channel space,
\begin{align}
t_1&=\left(\begin{array}{cc}{(t_1)}_{00}&{(t_1)}_{01}\\{(t_1)}_{10}&{(t_1)}_{11}\end{array}\right),~\quad t_2=\left(\begin{array}{cc}{(t_2)}_{00}&{(t_2)}_{01}\\{(t_2)}_{10}&{(t_2)}_{11}\end{array}\right),\nonumber\\
G_0&=\left(\begin{array}{cc}(G_0)_{00}&0\\0&(G_0)_{11}\end{array}\right),\label{input}
\end{align}
where for a given isospin $I$
\begin{align}
\mathcal{O}_{ab}&=\langle N(\mathbb{C}_1\mathbb{C}_2)_{I_b};I,I_z|\mathcal{O}|N(\mathbb{C}_1\mathbb{C}_2)_{I_a};I,I_z\rangle,\label{O}
\end{align}
with $\mathcal{O}=t_1$, $t_2$, $G_0$ and $\mathbb{C}_1$, $\mathbb{C}_2$ being the two particles forming a cluster with a defined isospin. To see how the elements of Eq.~(\ref{O}) can be determined, let us consider, for example, the $N(D\bar D^*)_0$ system. We can write,
\begin{align}
&|N(D \bar D^*)_0;I=1/2,I_z=1/2\rangle\nonumber\\
&\quad=|N;I_N=1/2,I_{zN}=1/2\rangle\nonumber\\
&\quad\quad\otimes |D \bar D^*; I_{D \bar D^*}=0,I_{z D\bar D^*}=0\rangle.\label{I1}
\end{align}
Since $t_1$ ($t_2$) is related to the $ND$ ($N\bar D^*$) interaction, to evaluate $(t_1)_{00}$ from Eq.~(\ref{O}) it is convenient to write Eq.~(\ref{I1}) in terms of the isospin of the $ND$ system, while to determine $(t_2)_{00}$ it is useful to express Eq.~(\ref{I1}) in terms of the isospin of the $N\bar D^*$ system. For the former case, using Clebsch-Gordan coefficients,  we get
\begin{align}
&|N(D \bar D^*)_0;I=1/2,I_z=1/2\rangle\nonumber\\
&\quad=\frac{1}{\sqrt{2}}\Big[|I_{ND}=1,I_{zND}=1\rangle\otimes \Big|I_{\bar D^*}=\frac{1}{2},I_{z\bar D^*}=-\frac{1}{2}\Big\rangle\nonumber\\
&\quad\quad-\frac{1}{\sqrt{2}}\Big(|I_{ND}=1,I_{zND}=0\rangle+|I_{ND}=0,I_{zND}=0\rangle\Big)\nonumber\\
&\quad\quad\otimes\Big|I_{\bar D^*}=\frac{1}{2},I_{z\bar D^*}=\frac{1}{2}\Big\rangle\Big],
\end{align}
and using now Eq.~(\ref{O}),

\begin{align}
&(t_1)_{00}=\langle N(D \bar D^*)_0;I=1/2,I_z=1/2|t_1\nonumber\\
&\quad\quad\quad\quad\times|N(D \bar D^*)_0;I=1/2,I_z=1/2\rangle\nonumber\\
&\quad=\frac{1}{4}(3 t^{1}_{ND}+t^{0}_{ND}).\label{t100}
\end{align}

In Eq.~(\ref{t100}), $t^0_{ND}$ ($t^1_{ND}$) corresponds to the two-body $t$-matrix for the $ND\to ND$ transition in isospin 0 (isospin 1). This process has been investigated within different models by solving the Bethe-Salpeter equation in a coupled channel approach (see, for example, Refs.~\cite{Hofmann:2005sw,Mizutani:2006vq,GarciaRecio:2008dp,Romanets:2012hm,Liang:2014kra}). All studies point to a common finding: the $DN$, $\pi\Sigma_c$ and coupled channel dynamics is attractive and gives rise to the formation of $\Lambda_c(2595)$~($J^P=1/2^-$). In Refs.~\cite{Hofmann:2005sw,Mizutani:2006vq}, the pseudoscalar-nucleon interactions with charm $+1$ are deduced from a Lagrangian based on the SU(4) symmetry. In Refs.~\cite{GarciaRecio:2008dp,Romanets:2012hm}, by using a model based on the SU(8) spin-flavor symmetry, which is compatible with the heavy-quark symmetry, pseudoscalar-baryon as well as vector-baryon channels are considered as coupled systems and generation of several $J^P=1/2^-$, $3/2^-$ $\Lambda_c$ and $\Sigma_c$  states is reported. In the latter references, $\Lambda_c(2595)$ is found to have large couplings to the $DN$ as well as to the $D^*N$ channel. The studies in Refs.~\cite{GarciaRecio:2008dp,Romanets:2012hm} have been updated more recently in Ref.~\cite{Nieves:2019nol} for the $\Lambda_c$ sector and it is again concluded that $\Lambda_c(2595)$ has a predominant molecular nature. However, it is suggested that $\Lambda_c(2625)$ ($J^P=3/2^-$) should be viewed mostly as a dressed three-quark state. In Ref.~\cite{Liang:2014kra}, within a different formalism based on arguments of heavy-quark and SU(4) symmetries, the $DN$, $\pi\Sigma_c$ and other coupled pseudoscalar-baryon channels and the $D^*N$, $\rho\Sigma_c$, and other coupled vector-baryon channels systems are studied. In this latter work, box diagrams are considered to construct a transition amplitude for the process $DN\leftrightarrow D^*N$. Several $\Lambda_c$ and $\Sigma_c$ states with $J^P=1/2^-$ and $3/2^-$, including $\Lambda_c(2595)$, are found. A large coupling of $\Lambda_c(2595)$ to $DN$ and $D^*N$ is also obtained as in Refs.~\cite{GarciaRecio:2008dp,Romanets:2012hm}. In the present work, we have considered the models of Refs.~\cite{GarciaRecio:2008dp,Liang:2014kra}, since in both cases the $DN$ and $D^*N$ channels are coupled, which is compatible with the heavy-quark symmetry. However, since the $\Lambda_c$ and $\Sigma_c$ states obtained in Refs.~\cite{GarciaRecio:2008dp,Liang:2014kra} are not all same, and many of them have not been observed experimentally yet, when investigating the $ND\bar D^*$ ($N\bar D D^*$) system, we focus mainly on the energy region in which $\Lambda_c(2595)$ is generated, since all models find similar properties and attribute a molecular nature to it. 

Coming back to the discussions on the three-body formalism, proceeding in a similar way as in Eq.~(\ref{t100}), we can get the rest of the elements in Eq.~(\ref{input}), for which we need the two-body $t$-matrices for the $N\bar D$ and $N\bar D^*$ systems in isospins 0 and 1. Within the SU(8) model of Refs.~\cite{GarciaRecio:2008dp,Romanets:2012hm},  in Ref.~\cite{Gamermann:2010zz} the $N\bar D$ and $N\bar D^*$ coupled system has been studied and several $\Lambda_{\bar c}$ and $\Sigma_{\bar c}$ with $J^P=1/2^-$ and $3/2^-$ have been claimed to get generated with masses in the energy region $2800-3000$ MeV. But, so far, no clear experimental evidence on the existence of such states is available, although the existence of an anticharm baryon with mass around 3000 MeV was claimed in Ref.~\cite{Aktas:2004qf}. However, subsequent experimental investigations have failed to confirm the former finding~\cite{Aubert:2006qu}. Within the model of Ref.~\cite{Mizutani:2006vq}, where $\bar D N$ and $\bar D^* N$ are not considered as coupled channels, we do not find formation of any state since the interaction is repulsive for the charm $-1$ sector. If we extend the model of Ref.~\cite{Liang:2014kra} to the charm $-1$ sector, we find formation of a $\Lambda_{\bar c}$ with a mass around 2950 MeV. Due to the different predictions within the models of Refs.~\cite{Mizutani:2006vq,Liang:2014kra,Gamermann:2010zz}, and the absence of conclusive experimental evidences in favor of the existence of any states with charm $-1$, we adopt the same strategy as in the charm $+1$ sector. We thus consider $\bar DN$ and $\bar D^* N$ as coupled channels within the models of Refs.~\cite{Gamermann:2010zz,Liang:2014kra} and restrict ourselves to an energy region in which the findings of the models, including the one of Ref.~\cite{Mizutani:2006vq}, are compatible.  In this way, our predictions for three-body states would be consistent with different input two-body amplitudes. 

Proceeding further with the discussion of the FCA, in Eq.~(\ref{input}), the propagator $G_0$ is given by~\cite{Xie:2010ig}
\begin{align}
(G_0)_{aa}=\frac{1}{2M_a}\int\frac{d^3q}{(2\pi)^3}\frac{m_N}{\omega_N(\vec{q})}\frac{F_a(\vec{q})}{q^0-\omega_N(\vec{q})+i\epsilon},\label{G0aa}
\end{align}
with $M_a$ being the mass of the cluster formed by $D\bar D^*$ ($\bar D D^*$), $m_N$ [$\omega_N(\vec{q})=\sqrt{\vec{q}^{\,2}+m^2_N}$] represents the mass (energy) of the nucleon which rescatters off the components of the cluster and $q^0$ is the nucleon on-shell energy in the nucleon-cluster rest frame, i.e.,
\begin{align}
q^0=\frac{s+m^2_N-M^2_a}{2\sqrt{s}},
\end{align}
where $\sqrt{s}$ is the center-of-mass energy of the three-body system. The function $F_a(\vec{q})$ in Eq.~(\ref{G0aa}) is a form factor related to the wave function of the particles of the cluster and it is given by~\cite{MartinezTorres:2020hus,MartinezTorres:2010ax}
\begin{align}
F_a(\vec{q})&=\frac{1}{\mathbb{N}}\int\limits_{|\vec{p}|,|\vec{p}-\vec{q}|<\Lambda}d^3pf_a(\vec{p})f_a(\vec{p}-\vec{q}),\nonumber\\
f_a(\vec{p})&=\frac{1}{\omega_{a1}(\vec{p})\omega_{a2}(\vec{p})}\frac{1}{M_a-\omega_{a1}(\vec{p})-\omega_{a2}(\vec{p})+i\epsilon},\label{FF}
\end{align}
where $\omega_{a1(a2)}(\vec{p})=\sqrt{\vec{p}+M^2_{a1(a2)}}$ are the energies of the particles in the cluster  and $\mathbb{N}$ is a normalization factor, $\mathbb{N}=F_a(\vec{q}=0)$. The upper limit of the integration in Eq.~(\ref{FF}) is related to the cut-off used when regularizing the two-body loops in the Bethe-Salpeter equation to generate $X(3872)$ or $Z_c(3900)$ from the coupled channel interactions. We consider a value for $\Lambda\sim 700$ MeV as in Refs.~\cite{Aceti:2014uea,Gamermann:2006nm,Aceti:2012cb} and vary it in a reasonable region to determine the uncertainties involved in the results. The unstable character of $Z_c(3900)$ is implemented in the formalism by substituting $M_a\to M_a-i\frac{\Gamma_a}{2}$ in Eq.~(\ref{FF}) (a width of 28 MeV is considered in this case). Note that the $1/(2M_a)$ present in Eq.~(\ref{G0aa}) is a normalization factor whose origin lies in the normalization of the fields when comparing the scattering matrix $S$ for a process in which a particle, in this case, a nucleon, rescatters off particles $\mathbb{C}_1$ and $\mathbb{C}_2$ of a cluster $\mathbb{C}$ and the $S$-matrix for an effective two-body particle-cluster scattering~\cite{MartinezTorres:2020hus,Ren:2018pcd}. As a consequence of the normalization of those $S$-matrices, 
\begin{align}
(G_0)_{aa}\to \frac{1}{2M_a}(G_0)_{aa},
\end{align}
and a normalization factor needs to be included in the two-body $t$-matrices $t_1$ and $t_2$ too,
\begin{align}
(t_1)_{ab}\to \sqrt{\frac{M_a}{M_{1a}}}\sqrt{\frac{M_b}{M_{1b}}}(t_1)_{ab},\\
(t_2)_{ab}\to \sqrt{\frac{M_a}{M_{2a}}}\sqrt{\frac{M_b}{M_{2b}}}(t_2)_{ab}.
\end{align}

With these ingredients we first solve Eq.~(\ref{Tpart}) and determine the $T$-matrix from Eq.~(\ref{T}) as a function of the center-of-mass energy $\sqrt{s}$ for the $N(D\bar D^*)$ and $N(\bar D D^*)$ systems. We then construct the $T$-matrix of Eq.~(\ref{NCa}) and search for peaks in $|T|^2$, which are identified with three-body states generated from the $NX(3872)/NZ_c(3900)$ coupled channel dynamics. 

\section{Results}
\subsection{Isospin 1/2}
\begin{figure}[h!]
\centering
\includegraphics[width=0.3\textwidth,height=5.5cm]{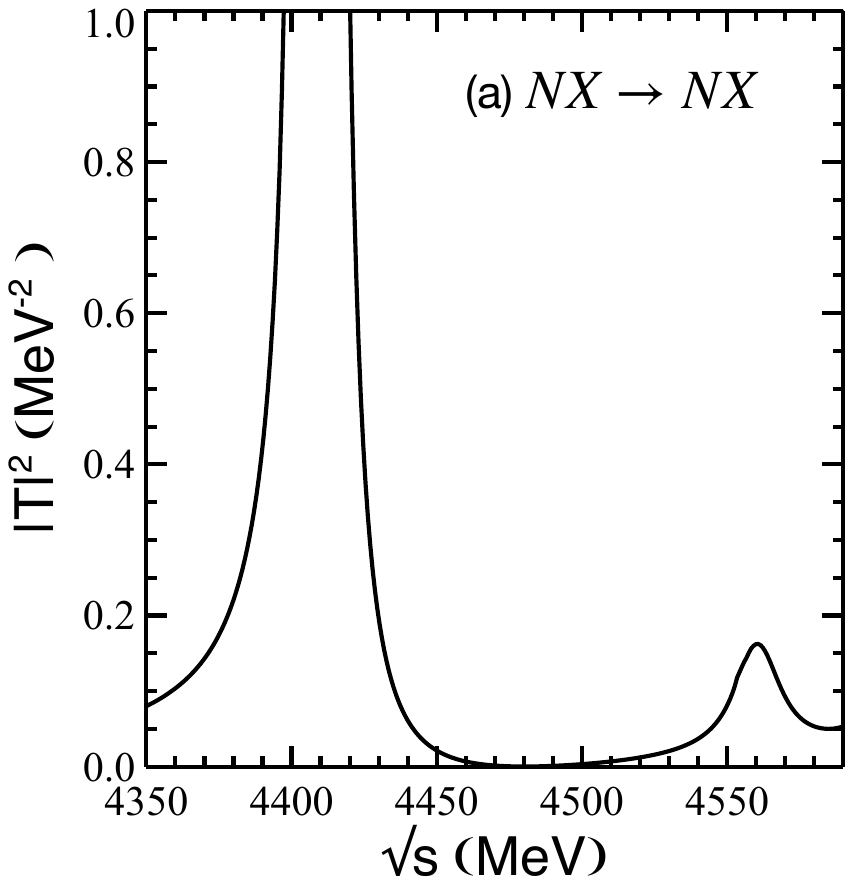}\vspace{0.4cm}
\includegraphics[width=0.3\textwidth,height=5.5cm]{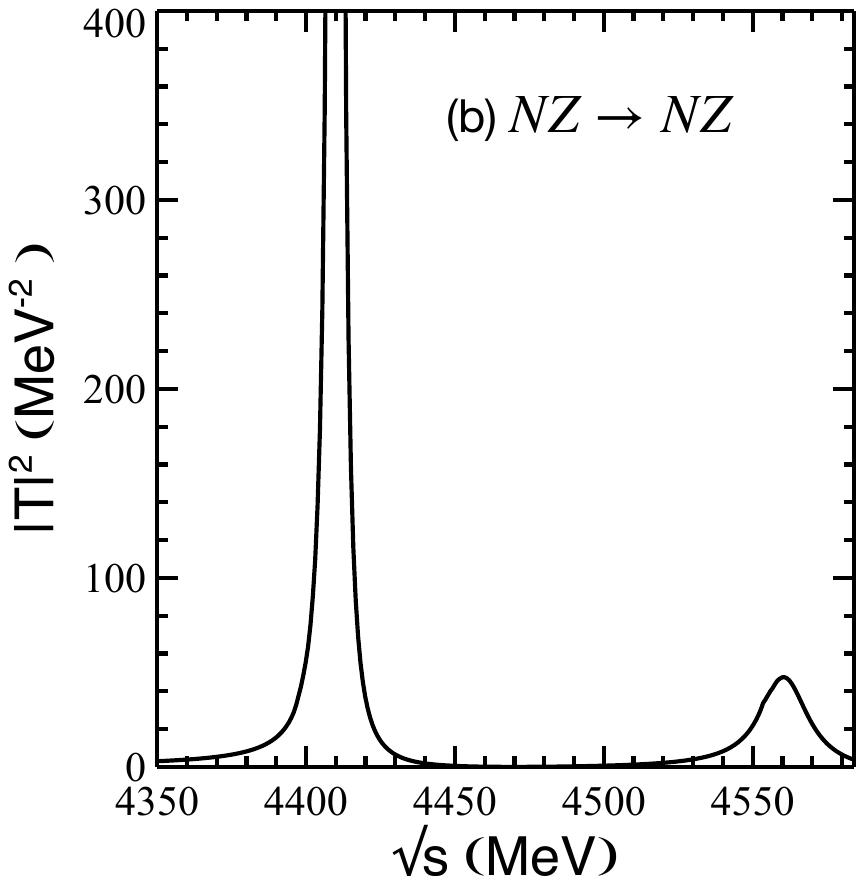}
\caption{Modulus squared of the $T$-matrix for the (a) $NX\to NX$  and (b) $NZ_c\to NZ_c$  transitions for $I(J^P)=1/2~(1/2^+)$ as a function of $\sqrt{s}$.}\label{Nstar_hidden_1h_1h}
\end{figure}

In Fig.~\ref{Nstar_hidden_1h_1h} we show the $|T|^2$ for isospin $1/2$ and $J^P=1/2^+$ for the (a) $NX\to NX$  and (b) $NZ_c\to NZ_c$ transitions obtained by using the two-body amplitudes determined from the model based on the heavy-quark and SU(4) symmetries~\cite{Liang:2014kra}. As can be seen, the three-body dynamics generates two states at $M-i\frac{\Gamma}{2}=4410-i1$ MeV and $4560-i10$ MeV, respectively, where $M$ and $\Gamma$ represent the mass and width found. The results shown in Fig.~\ref{Nstar_hidden_1h_1h} correspond to a value of $\Lambda=700$ MeV in Eq.~(\ref{FF}). We find that changing $\Lambda$ in a reasonable range, 700-770 MeV, produces small shifts in the masses of the states, $\sim 3-5$ MeV. Similar results are found if we determine the two-body amplitudes from the SU(8) Lagrangian of Ref.~\cite{GarciaRecio:2008dp}, with the corresponding peak positions being $4404-i1$ MeV and $4556-i2$ MeV. The results obtained from different two-body interactions and different cut-offs used in the form factor provide us with an estimate of the uncertainties present in the model.

\begin{figure}[h!]
\centering
\includegraphics[width=0.25\textwidth]{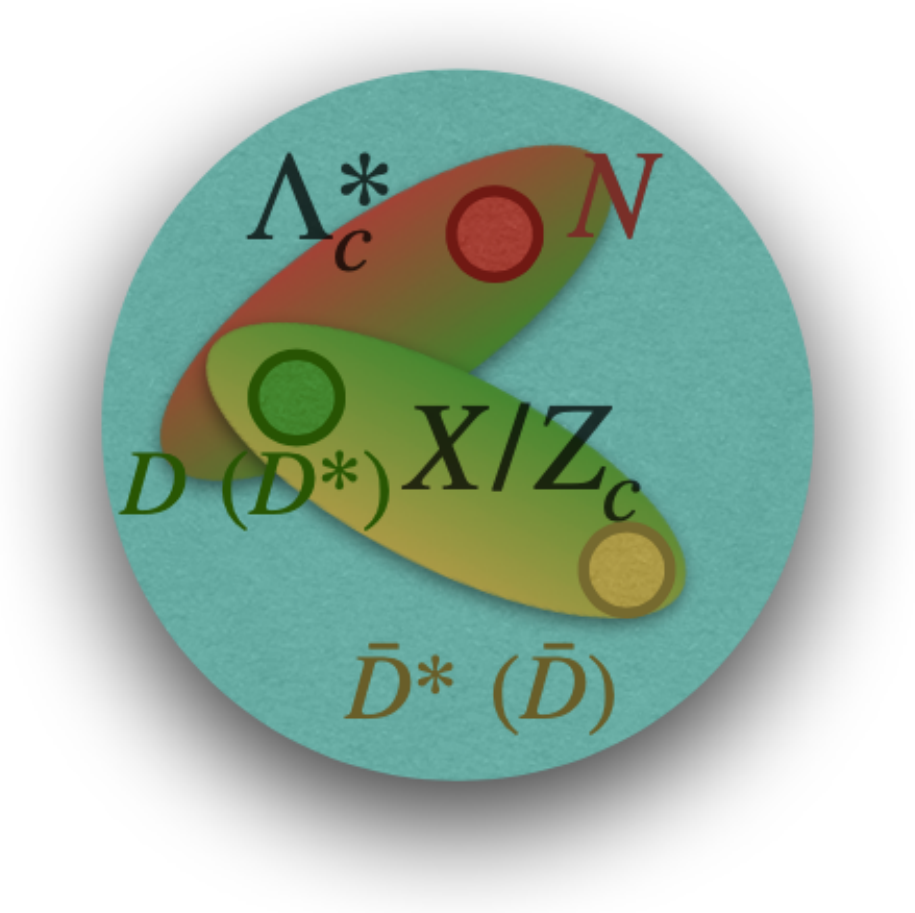}
\caption{Internal structure of the $N^*$ states with hidden charm obtained. The $ND-ND^*$ interaction generates $\Lambda_c (2595)$ while $D\bar D^*-\bar D D^*$ clusters as $X(3872)$/$Z_c(3900)$. In the energy range considered, the $N\bar D-N\bar D^*$ interaction does not give rise to any state.}\label{Nhid}
\end{figure}

The states obtained correspond to $N^*$'s with hidden charm and are generated when the $DN-D^*N$ subsystem forms $\Lambda_c(2595)$ while $D\bar D^*-\bar D D^*$ clusters as $X(3872)$ or $Z_c(3900)$ (see Fig.~\ref{Nhid}). In this sense, the corresponding wave functions would have an overlap with molecular $\Lambda_c(2595)\bar D^{(*)}$ components, besides $NX-NZ_c$. 
The formation of states from the interaction of a $\Lambda_c(2595)$ and a $\bar D$ has been suggested in Refs.~\cite{Burns:2019iih,Burns:2015dwa,Geng:2017hxc}. In particular,  the possibility of interpreting  the $P_c$ states observed in Refs.~\cite{Aaij:2015tga,Aaij:2016phn,Aaij:2019vzc} as $\Sigma_c\bar D^*-\Lambda_c(2595)\bar D$ molecules was discussed in the former works. In Ref.~\cite{Burns:2015dwa}, such a possibility was claimed based on an analogy between the mass difference of $D^*$ and $D$, with $X(3872)$ being formed in the $D\bar D^*+\text{c.c.}$ system, and the mass difference of $\Lambda_c(2595)$ and $\Sigma_c$. In this way, according to Ref.~\cite{Burns:2015dwa}, the $\Sigma_c\bar D^*-\Lambda_c(2595)\bar D$ system could generate a bound state in analogy with $X(3872)$ formed in $D\bar D^*+\text{c.c}$.  In Ref.~\cite{Geng:2017hxc}, by using scale invariance and arguments of heavy quark symmetry, the $\Sigma_c\bar D^*-\Lambda_c(2595)\bar D$ potential was obtained by means of pion exchange. It was found that for $J^P=1/2^+$, the attraction in the system is strong enough to generate bound states, pointing in this way to the existence of $1/2^+$ $\Sigma_c\bar D^*-\Lambda_c(2595)\bar D$ molecules, but no detailed calculation of the expected mass of such states was presented. Since the nominal mass of $P^+_c(4457)$ is close to the $\Sigma_c\bar D^*-\Lambda_c(2595)\bar D$ thresholds, the former could be a candidate for such a kind of molecular state. This suggestion was further considered in Ref.~\cite{Burns:2019iih}, where the potentials for the inelastic $\Sigma_c\bar D^*-\Lambda_c(2595)\bar D$ as well as for the elastic $\Sigma_c\bar D^*-\Sigma_c\bar D^*$ channels were obtained by using arguments of heavy quark symmetry in the former case, as in Ref.~\cite{Geng:2017hxc}, and the quark model in the latter. By varying the parameters of the model, the authors of Ref.~\cite{Burns:2019iih} find bound states with $J^P=1/2^+$ and $3/2^-$, with the former having a larger mass than the latter. In view of the proximity of $P^+_c(4457)$ to the $\Sigma_c\bar D^*$ and $\Lambda_c(2595)\bar D$ thresholds, the parameters of the model were adjusted to set the mass of the $J^P=3/2^-$ state to the nominal mass of $P^+_c(4440)$ and the $1/2^+$ state found was identified with $P^+_c(4457)$. 

In our formalism, although the wave function of the states obtained have an overlap with $\Lambda_c(2595)\bar D$, the dynamics considered is different to that studied in Refs.~\cite{Burns:2019iih,Geng:2017hxc}. In our case, $\Lambda_c(2595)$ is generated from the interaction of $DN$, $D^*N$ and other coupled channels, and $X(3872)/Z_c(3900)$ are treated as clusters of the $D\bar D^*-\bar D D^*$ system. In this way, we explicitly consider the three-body dynamics involved in the $ND\bar D^*-N\bar D D^*$ system, instead of treating it as an effective two-body system. We also consider all the interactions in s-wave. We must mention that since the cut-offs present in the calculation of the input two-body $t$-matrices, which are also used to set the value of $\Lambda$ in Eq.~(\ref{FF}), are fixed to reproduce the properties of $\Lambda_c(2595)$ and $X(3872)/Z_c(3900)$, respectively, there are no parameters in our model which could produce a significant shift of the masses obtained for the $N^*$ states.

\begin{figure}[h!]
\centering
\includegraphics[width=0.3\textwidth,height=5.5cm]{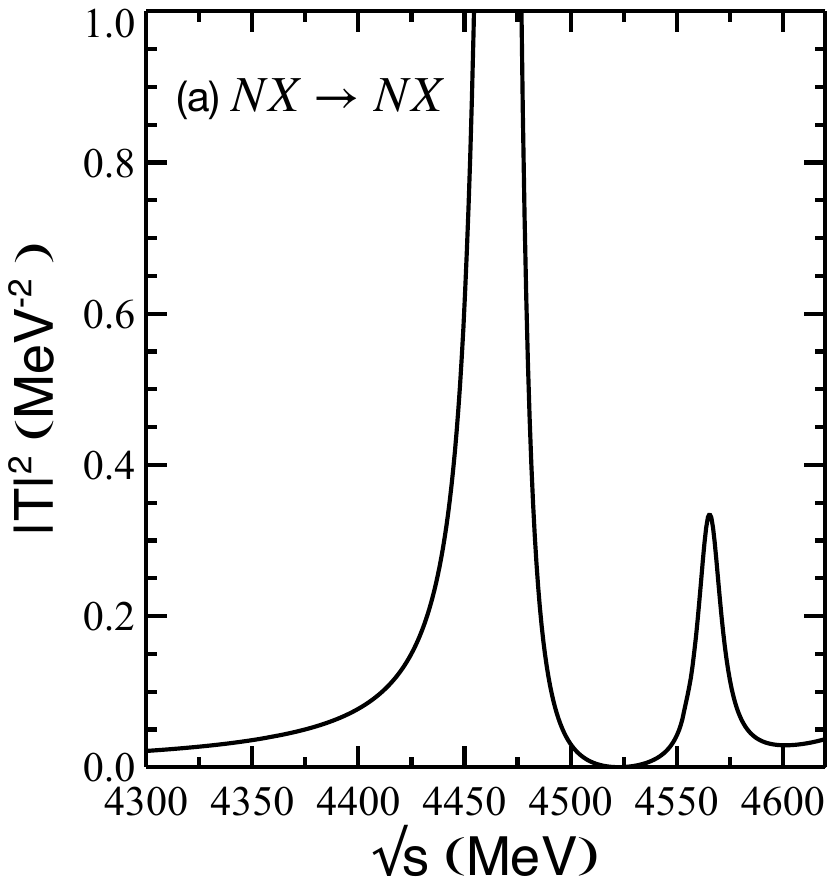}\vspace{0.4cm}
\includegraphics[width=0.3\textwidth,height=5.5cm]{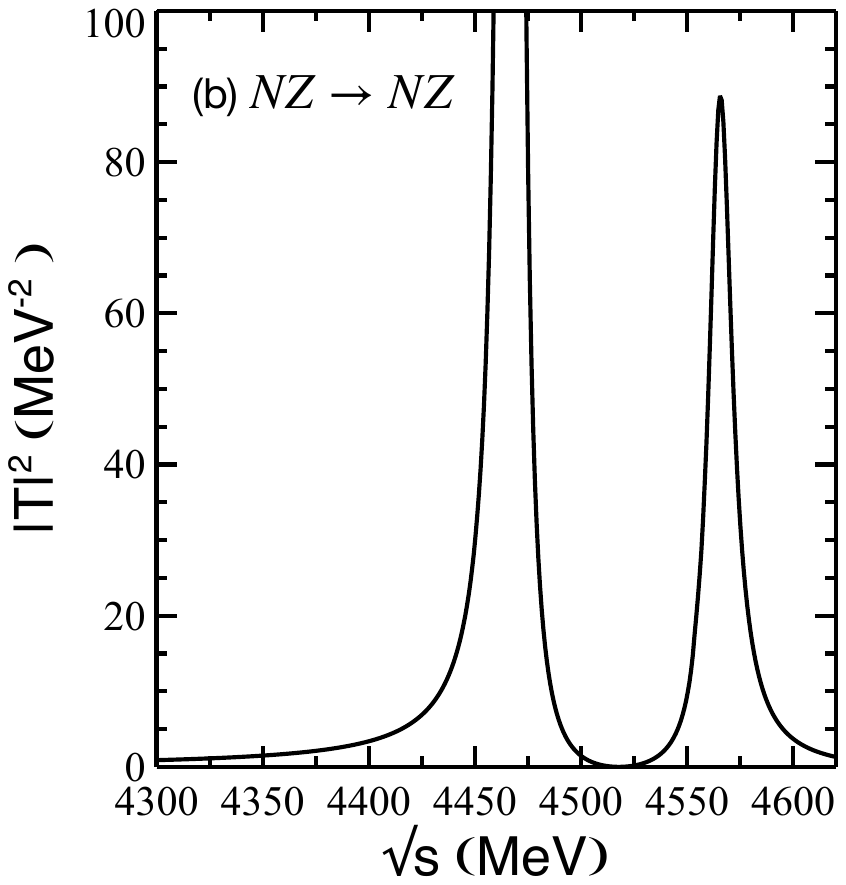}
\caption{Modulus squared of the $T$-matrix for the (a) $NX\to NX$  and (b) $NZ_c\to NZ_c$  transitions for $I(J^P)=1/2~(3/2^+)$ as a function of $\sqrt{s}$.}\label{Nstar_hidden_1h_3h}
\end{figure}

Considering only the formation of $\Lambda_c(2595)$ in the input $DN-D^*N$ two-body $t$-matrices, no state is obtained with $J^P=3/2^+$. However, meson-baryon interactions with charm $+1$ are found to be attractive at higher energies as well, within different approaches, leading to generation of further states. For example, in Refs.~\cite{GarciaRecio:2008dp,Liang:2014kra} a $\Lambda_c$ with $J^P=1/2^-$ and another with $J^P=3/2^-$, both with molecular nature and masses in the energy region $2600-2660$ MeV, are found, although the masses and widths obtained within the two approaches do not coincide and no experimental evidence for these states has been found yet. Nevertheless, if we consider the generation of these $\Lambda_c$'s in the input $DN-D^*N$ two-body $t$-matrices, three-body states with $J^P=3/2^+$ can be obtained. In Fig.~\ref{Nstar_hidden_1h_3h} we show the modulus squared of the $T$-matrix for the (a) $NX\to NX$  and (b) $NZ_c\to NZ_c$ transitions in isospin $1/2$ and $J^P=3/2^+$ determined following the model of Ref.~\cite{Liang:2014kra}. As can be seen, formation of two states, one at $4467-i3$ MeV and another at $4565-i7$ MeV, is found. In case of using the SU(8) model of Ref.~\cite{GarciaRecio:2008dp} for the two-body amplitudes, the corresponding peak positions are $4513-i1.3$ MeV and $4558-i2.4$ MeV.

\begin{figure}[h!]
\centering
\includegraphics[width=0.45\textwidth]{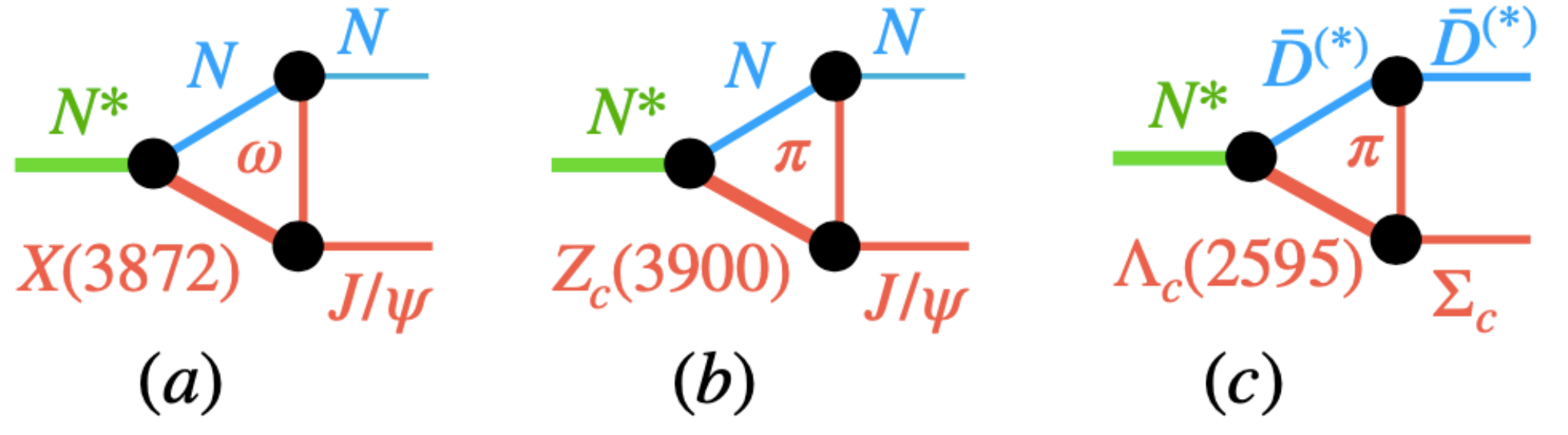}
\caption{Some of the decay modes of the $N^*$ states with hidden charm found in this work to two-body channels.}\label{decay}
\end{figure}

The $N^*$ states generated from the $ND\bar D^*-N\bar D D^*$ interactions, due to their three-body nature, can decay to three-body channels like $NJ/\psi\gamma$ and $NJ/\psi\pi$. Due to the formation of $\Lambda_c$ states in the $DN-D^*N$ subsystems when generating the $N^*$ states, decay channels like $\pi\Sigma_c\bar D$ can also be useful to study the properties of these states. Note that two-body decay channels like $J/\psi p$, $\bar D^{(*)}\Sigma_c$ can also exist, involving in this case triangular loops (see Fig.~\ref{decay}). Although the strength of the signal in such two-body invariant masses might not be large enough for a clear detection of the states. Indeed, the $J/\psi p$ invariant mass distribution reconstructed in Ref.~\cite{Aaij:2019vzc} shows fluctuations around 4400 MeV and 4550 MeV which could correspond to some of the states obtained in this work, and data with higher statistics would be necessary for confirming it. Interestingly, the $I(J^P)=1/2~(3/2^+)$ state obtained at $4467-i3$ MeV is compatible with the mass  and width of the $P^+_c(4457)$ claimed in Ref.~\cite{Aaij:2019vzc}.  

Further, we would like to add that the available data on weak decay processes like $\Lambda_b \to X(3872) p K^-$~\cite{Aaij:2019zkm} and $\Lambda_b \to  p J/\Psi \pi^+ \pi^- K^-$~\cite{Aaij:2016wxd} can be analyzed to investigate the states found in our work. In the former case, the invariant mass spectrum of $p K^-$ has been reconstructed and it shows the signal of $\Lambda(1520)$. The reconstruction of the $X(3872) p$ invariant mass can be useful in finding the nucleon states predicted in our work. The reconstruction of the invariant mass spectrum of $p J/\Psi \pi^-$ using data from Ref.~\cite{Aaij:2016wxd} can also show hidden charm states with positive parities. The states predicted in this work can also be studied at FAIR, in processes like $\bar p d \to n J/\Psi \pi$, $n J/\Psi \pi \pi$, $\bar D \Sigma_c \pi$, etc.

\subsection{Isospin 3/2}
So far we have restricted the discussions to the formation of $N^*$'s, where the dominant two-body interactions form well known resonances and different models agree on the description of such interactions. For example, there exists a general agreement on the strong coupling or association of the  $D^{(*)}\bar D^{(*)}$ isospin 0 and 1 interactions with $X(3872)$ and $Z_c(3900)$, respectively~\cite{Braaten:2003he,AlFiky:2005jd,Gamermann:2007fi,Gamermann:2010zz,Aceti:2014uea,Ortega:2018cnm,He:2017lhy}. Similarly the $ND^{(*)}$ isoscalar interaction is known to be attractive in nature, leading to the formation of $\Lambda_c(2595)$ in various models~\cite{Hofmann:2005sw,Mizutani:2006vq,GarciaRecio:2008dp,Romanets:2012hm,Liang:2014kra,Nieves:2019nol}. There exists enough experimental data to define the properties of  $X(3872)$,  $Z_c(3900)$ and $\Lambda_c(2595)$. The reliability of the description of the isovector $ND^{(*)}$ interaction within different models, however, is difficult to judge at this moment since $\Sigma_c$ states are not yet well known experimentally. For example, only three $\Sigma_c$ states are listed in Ref.~\cite{pdg} by the Particle Data Group, out of which the quantum numbers of only two are known. Additionally, different theoretical models predict a different spectra. For example, as mentioned earlier, we have considered the theoretical works of Refs.~\cite{GarciaRecio:2008dp,Liang:2014kra} to describe the $ND^{(*)}$ interactions. In the former work, relatively narrow $\Sigma_c$'s with $J^P=1/2^-,\,3/2^-$  were found with mass around 2600 MeV (which is within the range of invariant masses considered in our study). Such states have not been observed experimentally yet. In Ref.~\cite{Liang:2014kra}, the isospin one $ND^{(*)}$ and coupled channel  interactions are found to be attractive, but only a very broad state (width $\sim300$ MeV) is obtained around 2600 MeV.  Still, it can be worth exploring the formation of three-body states with isospin 3/2, which requires all the two-body subsystems to interact in isospin 1.  Such states will be like $\Delta$'s with hidden charm, which must also, presumably, exist in nature. 

\begin{figure}
\centering
\includegraphics[width=0.3\textwidth]{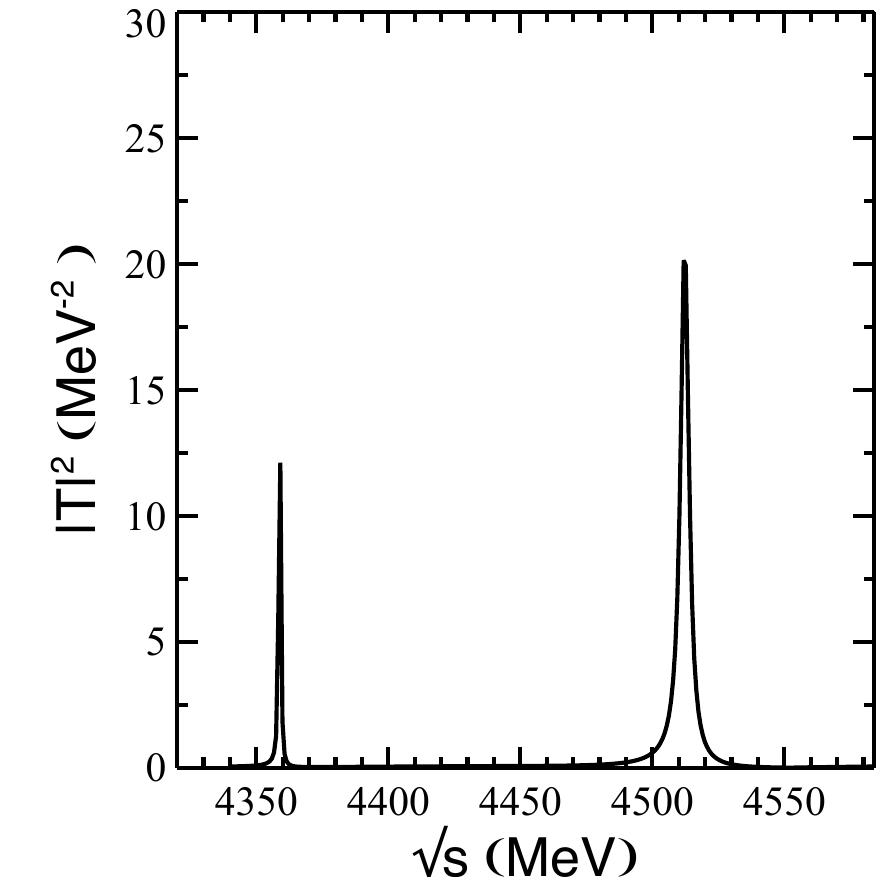}
\caption{Modulus squared of the $T$-matrix for $NZ_c\to NZ_c$  transitions with $I(J^P)=3/2~(1/2^+)$ as a function of $\sqrt{s}$.}\label{Deltastar}
\end{figure}

In Fig.~\ref{Deltastar}, we show the isospin 3/2 three-body modulus squared amplitude for total spin 1/2.  The amplitude depicted in Fig.~\ref{Deltastar} is computed with the $ND^{(*)}$ interactions which give rise to formation of  some $1/2^-, 3/2^-$ $\Sigma_c$'s in the energy region around $2600$ MeV~\cite{GarciaRecio:2008dp}.
It can be seen that two states arise from the interactions, one with mass around 4359 MeV and a width of about 1.5 MeV, another around 4512 MeV and width $\sim$4 MeV. Similar results are obtained  for total spin 3/2. We thus find almost spin degenerate states, which we denote as $\Delta_c$. 

We must add that within the model of Ref.~\cite{Liang:2014kra}, no such isospin 3/2 states are found. Thus, the existence of the $\Delta_c$'s shown in Fig.~\ref{Deltastar} is conditioned to the existence of narrow negative-parity  $\Sigma_c$ state(s)  with mass similar to $\Lambda_c(2595)$.

\section{Contributions from diagrams beyond the fixed center approximation}
\begin{figure}[h!]
\centering
\includegraphics[width=0.4\textwidth]{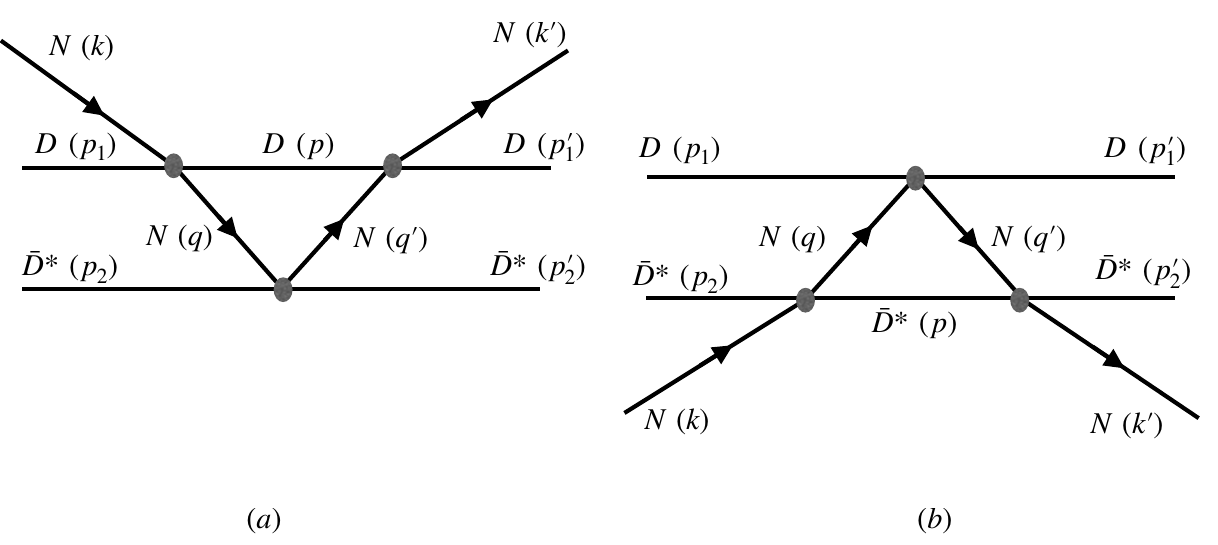}
\caption{One loop diagrams contributing to the two Faddeev series within the static approximation [see Eq.~(\ref{Tpart})]. The labels in the brackets represent the momenta assigned to each hadron.}\label{FCAdiagrams}
\end{figure}

\begin{figure}[h!]
\centering
\includegraphics[width=0.4\textwidth]{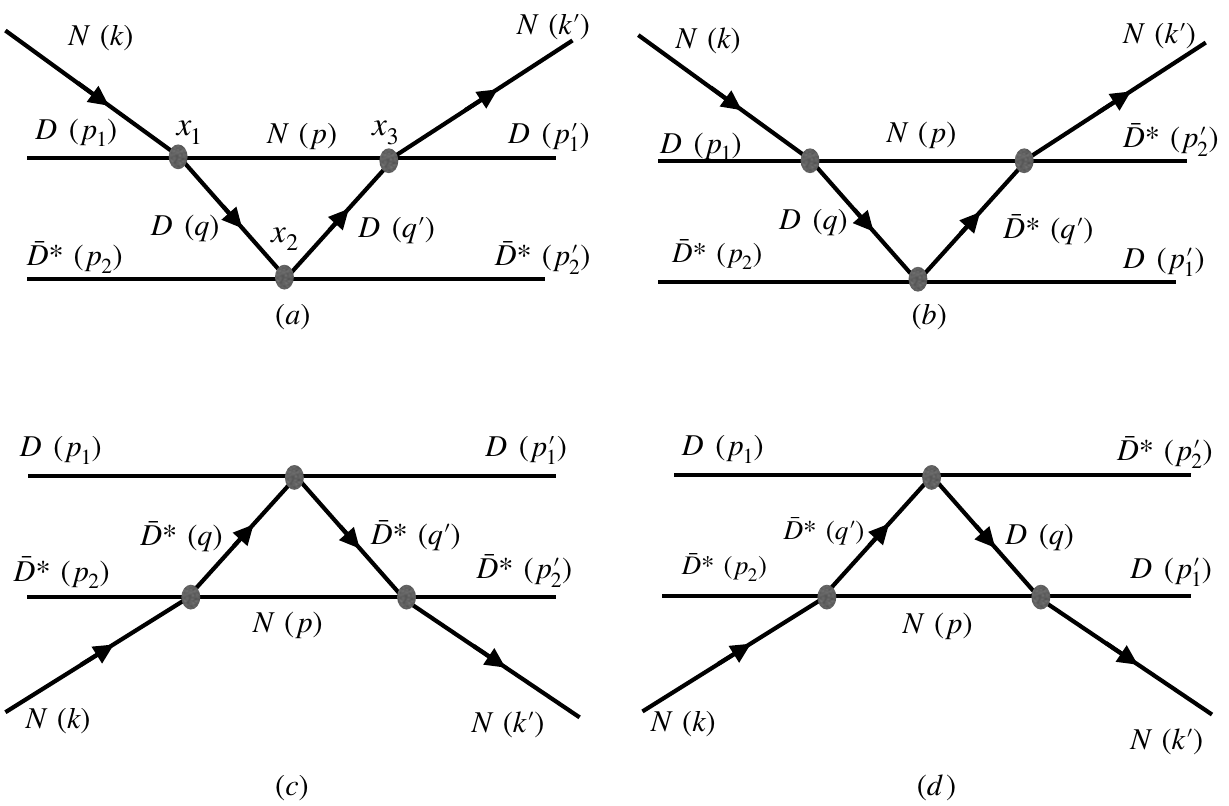}
\caption{Diagrams which can contribute at the one loop level, beyond the static approximation. The meaning of the labels in brackets is same as in Fig~\ref{FCAdiagrams}.  }\label{morediagrams}
\end{figure}

Before ending the discussions, we would like to analyze the uncertainties involved in our findings from additional diagrams beyond those involved in the FCA. In the current study, we consider that $D$ and $\bar D^*$ form a cluster which stays unperturbed during the scattering. The meaning of such a consideration is that the third hadrons scatters off the constituents of the cluster which act like static sources. Within such an approximation, at the level of one loop,  
the diagrams besides those shown in Fig.~\ref{FCAdiagrams} are considered to be suppressed. 
In the present case, where the mass of the nucleon is about half of the mass of  $D$/$\bar D^*$, one might wonder if the additional diagrams shown in Fig.~\ref{morediagrams} can be neglected and if the fixed center approximation is appropriate for the current system. To test the applicability of the approximation, we make explicit calculations of the diagrams in Fig.~\ref{morediagrams} in this section and show that the resulting contributions turn out to be negligible. We also discuss the reason behind such findings.

Before going to the details of the calculations,  we would like to remind the reader that the fixed center or the static approximation has been applied to studies of anti-kaon deuteron scattering in Refs.~\cite{Kamalov:2000iy,Chand:1962ec}, where the kaon is about half as heavy as the nucleon. Still, results on the scattering length were obtained in good agreement with the experimental data. Indeed contributions from recoil effects were scrutinized in detail in Ref.~\cite{Baru:2009tx} by considering a perturbative expansion in terms of the ratio of the masses $M_K/m_N$ and corrections of the order of 10-15$\%$ were found. Such a small correction was attributed to cancellations  among different terms in the perturbative series. The mentioned cancellations were attributed, in Ref.\cite{Baru:2009tx}, to the Pauli principle, or to the orthogonality of the deuteron wave function and the $NN$ continuum, depending on the isospin of the $\bar K N$ interactions. Similar conclusions were obtained in a later study in Ref.~\cite{Mai:2014uma} too. It should be mentioned that besides the anti-kaon deuteron case, cancellations have been found in the case of the $\pi d$ scattering too~\cite{Faldt:1974sm}, where even though the static approximation may be expected to work well, corrections (from binding energy) turn out to be large when considering each term of the scattering series separately. However, corrections to the different terms end up   canceling with each other when the series is summed~\cite{Faldt:1974sm}, rendering the FCA applicable to the system. Interestingly, validity of the FCA was also discussed in Ref.~\cite{MartinezTorres:2010ax} in case of the $\phi K\bar K$ system. It was found that the FCA amplitude is not reliable in the former case, as expected, except for energies below the cluster-particle threshold, thereby limiting the prospects of the excitation of the constituents of the cluster.  Thus, the static approximation has been found to work in a series of unexpected systems due to different reasons. 

It might also be useful to cite  examples of some three-hadron systems which have been studied by solving the Faddeev equations with and without the consideration of the static approximation for one of the subsystems. For example, the $NK\bar K$ system and coupled channels were studied in Refs.~\cite{MartinezTorres:2008kh,MartinezTorres:2010zv}  by solving the Faddeev equations without invoking the static approximation for any of the pairs. In this case a $1/2^+$ state, with mass around 1924 MeV, was found with the width ranging between 20-30 MeV. Similar results have been reported in Ref.~\cite{Jia:2011zzd}, where Faddeev equations were solved using an effective potential  for each of the pairs. A state with a mass around $1880-1920$ MeV is obtained in the former work. Further, the same system was studied by treating $\bar K N$ and $\bar K K$ as fixed scattering centers in Ref.~\cite{Xie:2010ig} and results compatible with those of Refs.~\cite{MartinezTorres:2008kh,MartinezTorres:2010zv,Jia:2011zzd} (mass $\sim1915-1925$ MeV, width $\sim 30-80$ MeV) were found. As shown in Ref.~\cite{MartinezTorres:2010ax}, the condition in which FCA seems to works well  is when a  three-body system is studied at energies below the threshold, besides having a two-body cluster which is heavier than the third one.  Yet another system, $DK\bar K$,  was studied by solving full Faddeev equations~\cite{MartinezTorres:2012jr} as well as by introducing the FCA~\cite{Debastiani:2017vhv}.  A $D$-meson with spin parity $1/2^-$, mass around 2900 MeV and with of 55 MeV was found to arise from the three-body interactions in Ref.~\cite{MartinezTorres:2012jr},  whose decay to two mesons has been studied in Ref.~\cite{Malabarba:2021gyq}.  Indeed a state with same quantum numbers was found in Ref.~\cite{Debastiani:2017vhv} but with the values of mass (and, consequently, width) about 50 MeV lower than those determined with full Faddeev calculations~\cite{MartinezTorres:2012jr}. However, it must be mentioned that the state found in Ref.~\cite{MartinezTorres:2012jr} appeared in the $Df_0(980)$ configuration, while a clear signal was not found in the $D_s(2317)\bar K$ arrangement of the three-body system. The latter is precisely the configuration studied in Ref.~\cite{Debastiani:2017vhv} in order to use the FCA (although the authors of Ref.~\cite{Debastiani:2017vhv} arrive to the conclusion that $Df_0(980)$ is the dominant configuration). From the above discussion one can see that, as far as the energy region studied is below the three-body threshold, the order of uncertainties  in the results obtained by using  FCA is similar to that found within other methods for solving the Faddeev equations where no static approximations are considered.

Let us now discuss the case of the $ND\bar D^*$ system by evaluating the amplitudes for the diagrams shown in Fig.~\ref{morediagrams}, where the $D$ and $\bar D^*$ can propagate as free particles in the loop. Following the same normalization as in Ref.~\cite{MartinezTorres:2010ax}, we can write the contribution to the $S$ matrix, to start with, for the diagram in Fig.~\ref{morediagrams}a as 
\begin{align}
S_{\ref{morediagrams}a}=&\sqrt{\frac{2M_N}{2k^0}}\sqrt{\frac{2M_N}{2k^{\prime 0}}}\sqrt{\frac{1}{2p_1^0}}\sqrt{\frac{1}{2p_2^0}}\sqrt{\frac{1}{2p_1^{\prime 0}}}\sqrt{\frac{1}{2p_2^{\prime 0}}}\nonumber\\
&\quad\times\int d^4x_1\int d^4 x_2\int d^4 x_3 \int \frac{d^4 q}{\left(2\pi\right)^4}\nonumber\\
&\quad\times\int \frac{d^4 q^\prime}{\left(2\pi\right)^4}\int \frac{d^4 p}{\left(2\pi\right)^4}\Bigl[-it_{DN}\left(k+p_1\right)\Bigr]\nonumber\\
&\quad\times\Bigl[-i t_{D\bar D^*}\left(\sqrt{s_{D\bar D^*}}\right)\Bigr]\Bigl[-it_{DN}\left(k^\prime+p_1^\prime\right)\Bigr]\nonumber\\
&\quad\times \frac{ie^{iq\left(x_1-x_2\right)}}{q^2-m_D^2+i\epsilon}\frac{ie^{iq^\prime\left(x_2-x_3\right)}}{q^{\prime 2}-m_{\bar D^*}^{2}+i\epsilon}\nonumber\\
&\quad\times\frac{ie^{ip\left(x_1-x_3\right)}\bar{u}(k^\prime)\left(\slashed{p}+m_N\right)u(k)}{p^2-m_N^2+i\epsilon}\nonumber\\
&\quad\times e^{ik^{\prime 0} x_3^0}e^{ip_1^{\prime 0} x_3^0}e^{ip_2^{\prime 0} x_2^0}e^{-ik^{0} x_1^0}e^{-ip_1^{0} x_1^0}e^{-ip_2^{0} x_2^0}\nonumber\\
&\quad\times\frac{1}{\sqrt{V}} e^{-i\vec k^\prime\cdot \vec x_3}\phi_1(x_3)\phi_2(x_2)\frac{1}{\sqrt{V}} e^{i\vec k\cdot \vec x_1}\phi_1(x_1)\phi_2(x_2),\label{s2a}
\end{align}
where $\phi_i (x_j)$ represent the wave functions of the particles of the cluster in the initial/final state. We refer the reader to Fig.~\ref{morediagrams}a to identify the momenta assigned to each particle. The invariant mass of the $D\bar D^*$ system, in Eq.~(\ref{s2a}), depends on a loop variable through 
\begin{align}
s_{D\bar D^*}=s+m_N^2-2\sqrt{s}~\omega_N(\vec p).
\end{align}
Integrating on the zero component of the six variables in Eq.~(\ref{s2a}) and defining 
\begin{align}
&\vec x_1-\vec x_2\equiv\vec r,\\\nonumber
 &\vec x_3-\vec x_2\equiv\vec r^{\,\prime}, \\\nonumber
&\vec R\equiv\frac{1}{2}\left(\vec x_1+\vec x_2\right),
\end{align}
we can write
\begin{align}
\frac{1}{2}\left(\vec x_3+\vec x_2\right)=\vec R+\frac{\vec r}{2}+\frac{\vec r^{\,\prime}}{2}.
\end{align}
Such change of variables allows us to write 
\begin{align}
\phi_1(x_1)\phi_2(x_2)=\frac{1}{\sqrt{V}}e^{i\vec P_{12} \cdot\vec R}\phi\left(\,\vec r\,\right),
\end{align}
and 
\begin{align}
\phi_1(x_3)\phi_2(x_2)=\frac{1}{\sqrt{V}}e^{-i\vec P^\prime_{12} \cdot\vec R}\phi\left(\,\vec r\,\right)e^{-i\vec P^\prime_{12} \cdot\vec r^{\,\prime}/2}e^{i\vec P^\prime_{12} \cdot\vec r/2}\phi\left(\vec {r}^{\,\prime}\right),
\end{align}
where $P_{12} \left(P^{\prime}_{12}\right)$ denotes the momentum of the cluster in the initial (final) state. Finally, integrating on $\vec r$, $\vec r^{\,\prime}$ and $\vec R$ we get 
\begin{align}\nonumber
&S_{\ref{morediagrams}a}=-i\frac{\left(2\pi\right)^4\delta^4\left(P_{tot}-P^\prime_{tot}\right)}{V^2}\sqrt{\frac{2M_N}{2k^0}}\sqrt{\frac{2M_N}{2k^{\prime 0}}}\nonumber\\
&\quad\times\sqrt{\frac{1}{16p_1^0p_2^0p_1^{\prime 0}p_2^{\prime 0}}} \sqrt{\frac{k^0+m_N}{2m_N}}\sqrt{\frac{k^{\prime 0}+m_N}{2m_N}}\nonumber\\
&\quad\times t_{DN}\left(k+p_1\right)t_{DN}\left(k^\prime+p_1^\prime\right)\int \frac{d^4 q}{\left(2\pi\right)^4}\int \frac{d^4 q^\prime}{\left(2\pi\right)^4}\nonumber\\
&\quad\times\int \frac{d^4 p}{\left(2\pi\right)^4}t_{D\bar D^*}\left(\sqrt{s_{D\bar D^*}}\right)\frac{\omega_N\left(\vec p\right)+m_N}{2\omega_N \left(\vec p\right)} \nonumber\\
&\quad\times \frac{1}{\left[k^0+p_1^0-\omega_N\left(\vec p\right)\right]^2-\omega\left(\vec q\right)^2+i\epsilon}\nonumber\\
&\quad\times\frac{1}{\left[k^{\prime 0}+p_1^{\prime 0}-\omega_N\left(\vec p\right)\right]^2-\omega\left(\vec q^{\,\prime} \right)^2+i\epsilon}\nonumber \\
&\quad\times \phi_1\left(\vec p +\vec q-\frac{\vec k}{2}\right)\phi_2\left(\vec p +\vec q^{\, \prime}-\frac{\vec k^{\,\prime}}{2}\right),\label{s2afin}
\end{align}
where $P_{tot}\left(P^\prime_{tot}\right)$ represents the total four momentum of the three body system in the initial (final) state and $\phi_1$, $\phi_2$ are calculated following Ref.~\cite{Gamermann:2009uq}. It must be emphasized here that the different two-body amplitudes get contributions from different isospin with different weights (as explained in section~\ref{Fo}). Following the same procedure, we can obtain the amplitudes for the remaining diagrams in Fig.~\ref{morediagrams} as
\begin{align}
&t_{\ref{morediagrams}b}=t_{\ref{morediagrams}d}= \sqrt{\frac{k^0+m_N}{2m_N}}\sqrt{\frac{k^{\prime 0}+m_N}{2m_N}} t_{DN}\left(k+p_1\right)\nonumber\\
&\quad\times t_{\bar D^*N}\left(k^\prime+p_2^\prime\right)\int \frac{d^4 q}{\left(2\pi\right)^4}\int \frac{d^4 q^\prime}{\left(2\pi\right)^4}\int \frac{d^4 p}{\left(2\pi\right)^4}\nonumber\\
&\quad\times\frac{\omega_N\left(\vec p\right)+m_N}{2\omega_N \left(\vec p\right)}\frac{t_{D\bar D^*}\left(\sqrt{s_{D\bar D^*}}\right)}{\left[k^0+p_1^0-\omega_N\left(\vec p\right)\right]^2-\omega\left(\vec q\right)^2+i\epsilon} \nonumber\\
&\quad\times \frac{1}{\left[k^{\prime 0}+p_2^{\prime 0}-\omega_N\left(\vec p\right)\right]^2-\omega\left(\vec q^{\,\prime} \right)^2+i\epsilon}\nonumber \\
&\quad\times \phi_1\left(\vec p +\vec q-\frac{\vec k}{2}\right)\phi_2\left(\vec p +\vec q^{\, \prime}-\frac{\vec k^{\,\prime}}{2}\right),\label{s2bfin}
\end{align}
and 
\begin{align}
&t_{\ref{morediagrams}c}=\sqrt{\frac{k^0+m_N}{2m_N}}\sqrt{\frac{k^{\prime 0}+m_N}{2m_N}} t_{\bar D^* N}\left(k+p_2\right)t_{\bar D^*N}\left(k^\prime+p_2^\prime\right)\nonumber\\
&\quad\times\int \frac{d^4 q}{\left(2\pi\right)^4}\int \frac{d^4 q^\prime}{\left(2\pi\right)^4}\int \frac{d^4 p}{\left(2\pi\right)^4}t_{D\bar D^*}\left(\sqrt{s_{D\bar D^*}}\right)\nonumber\\
&\quad\times\frac{\omega_N\left(\vec p\right)+m_N}{2\omega_N \left(\vec p\right)}\frac{1}{\left[k^0+p_2^0-\omega_N\left(\vec p\right)\right]^2-\omega\left(\vec q\right)^2+i\epsilon}\nonumber\\
&\quad\times  \frac{1}{\left[k^{\prime 0}+p_2^{\prime 0}-\omega_N\left(\vec p\right)\right]^2-\omega\left(\vec q^{\,\prime} \right)^2+i\epsilon} \\
&\quad\times \phi_1\left(\vec p +\vec q-\frac{\vec k}{2}\right)\phi_2\left(\vec p +\vec q^{\, \prime}-\frac{\vec k^{\,\prime}}{2}\right).\label{s2cfin}
\end{align}

\begin{figure}[h!]
\centering
\includegraphics[width=0.45\textwidth]{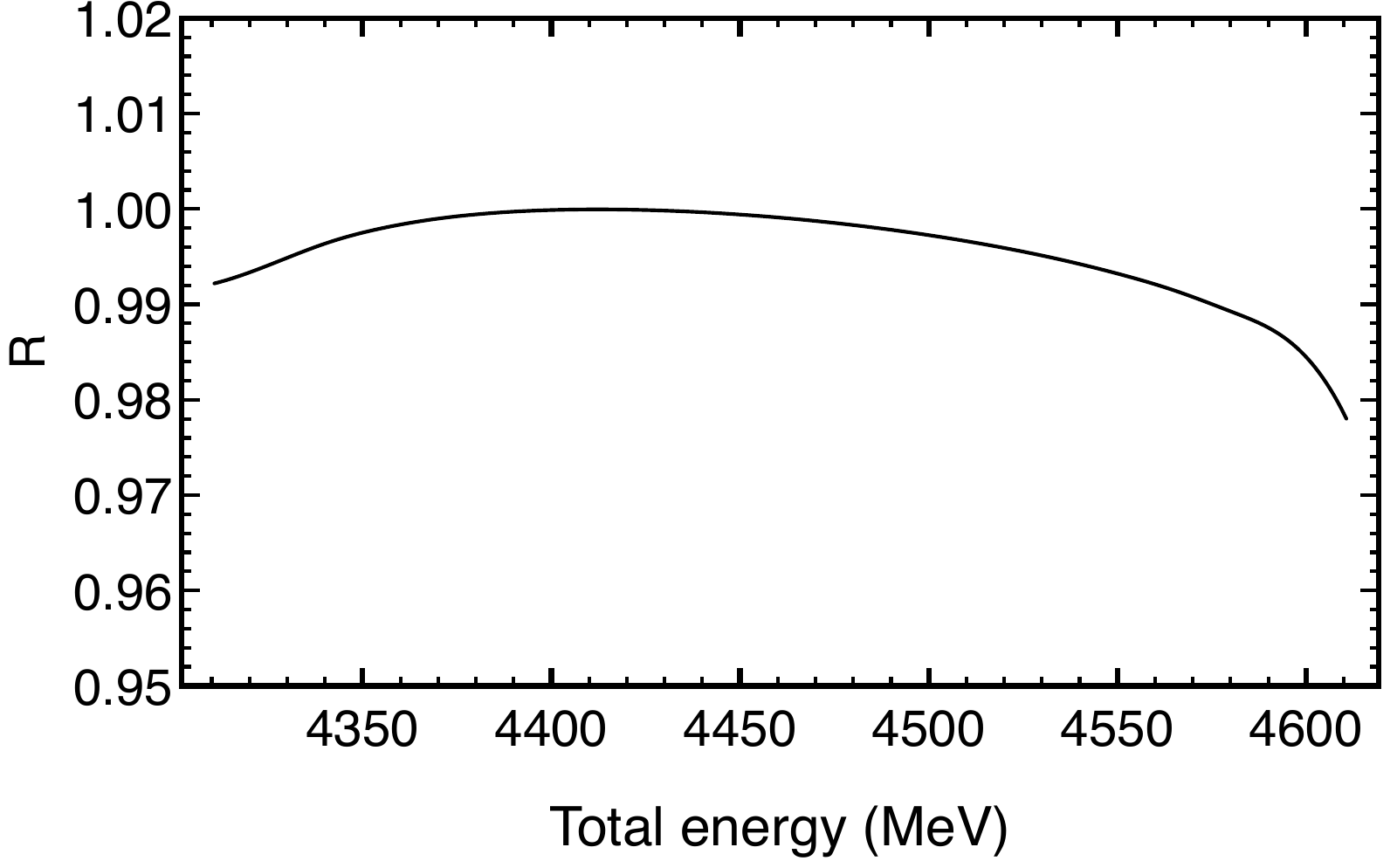}
\caption{The ratio defined in Eq.~(\ref{eqratio}) as a function of the total energy of the three-body system.}\label{ratio}
\end{figure}

Let us call the amplitude of the diagrams shown in Fig.~\ref{FCAdiagrams}, which contribute to the FCA series [Eqs.~(\ref{Tpart})], as $t_{\ref{FCAdiagrams}a}$ and $t_{\ref{FCAdiagrams}b}$. To study the effect of the considerations of the diagrams in Fig.~\ref{morediagrams}, which go beyond the FCA, we show in Fig.~\ref{ratio} the ratio 
\begin{equation}
R=\frac{|t_{\ref{FCAdiagrams}a}+t_{\ref{FCAdiagrams}b}+t_{\ref{morediagrams}a}+t_{\ref{morediagrams}b}+t_{\ref{morediagrams}c}+t_{\ref{morediagrams}d}|}{|t_{\ref{FCAdiagrams}a}+t_{\ref{FCAdiagrams}b}|}.\label{eqratio}
\end{equation}
It can be seen that the ratio stays very close to unity, showing that the contribution from the diagrams beyond the FCA provide a very small correction, and, hence, indicating the approximation to be indeed reliable in the present case.  

\begin{figure}[h!]
\centering
\includegraphics[width=0.45\textwidth]{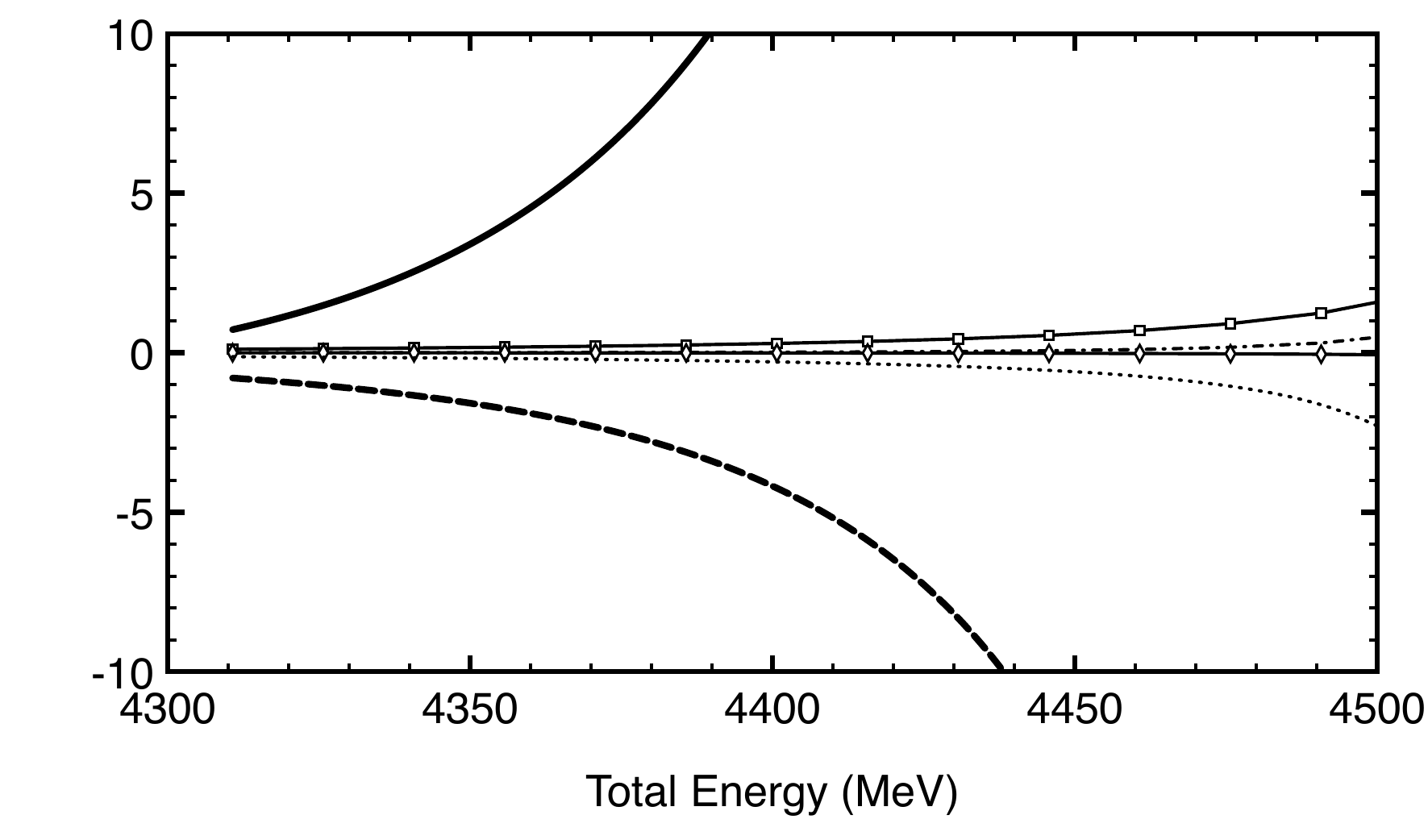}
\caption{Comparison of the different amplitudes for the diagrams shown in Figs.~\ref{FCAdiagrams} and \ref{morediagrams}. The thick solid (dashed) line shows the real (imaginary) part of the sum of the amplitudes for the one loop diagrams contributing to the FCA series, i.e., $t_{\ref{FCAdiagrams}a}+t_{\ref{FCAdiagrams}b}$. The dotted (dash-dotted) line represents the real (imaginary) part of the sum of the amplitudes for Figs.~\ref{morediagrams}a and \ref{morediagrams}c, while the line with boxes (line with diamonds) show the real (imaginary) part of the sum of  the amplitudes for Figs.~\ref{morediagrams}b and \ref{morediagrams}d. }\label{compare}
\end{figure}

To understand the reason behind such a small correction,  we compare the different amplitudes in Fig.~\ref{compare}. Firstly, it can be noticed that the sum of the amplitudes of the one-loop diagrams in the FCA series (see the real and imaginary parts represented as solid and dashed lines in Fig.~\ref{compare}) is much bigger than the amplitudes corresponding to the diagrams beyond the static approximation (see the caption of Fig.~\ref{compare} for more details). 
It should be mentioned that the limits on the vertical axis, in  Fig.~\ref{compare}, have been kept as shown in the figure to facilitate a comparison of the different amplitudes. Besides the small contributions from the diagrams in Fig.~\ref{morediagrams}, to which we will come back in a moment, it should be noticed that the sum of the amplitudes $t_{\ref{morediagrams}a}+t_{\ref{morediagrams}c}$ and  $t_{\ref{morediagrams}b}+t_{\ref{morediagrams}d}$ have opposite signs, leading to cancellations among the two. The real part of $t_{\ref{morediagrams}a}+t_{\ref{morediagrams}c}$ is shown as a dotted line, which should be compared with the real part of $t_{\ref{morediagrams}b}+t_{\ref{morediagrams}d}$, shown as a line with boxes. The imaginary parts of $t_{\ref{morediagrams}a}+t_{\ref{morediagrams}c}$ and $t_{\ref{morediagrams}b}+t_{\ref{morediagrams}d}$ are shown as a dash-dotted line and a line with diamonds, respectively. The imaginary parts are much smaller, which should be expected at energies below the threshold, though nonzero since there exist lighter (and, hence, open) coupled channels in the  two-body  subsystems. For example, $\pi \Lambda_c$ and $\pi \Sigma_c$ are open below the $DN$ threshold. In any case, we can question why the amplitudes for the  diagrams in Fig.~\ref{morediagrams} turn out to be much smaller. To understand this one must recall that: (1) We have more number of heavier particles ($D$/$\bar D^*$) propagating in the intermediate state in Fig.~\ref{morediagrams}, when compared to the diagrams shown in Fig.~\ref{FCAdiagrams}. (2) The energies of interest, where we find the states, are below the three-body threshold,  where the contributions from the excitation of the particles in the cluster are expected to be small (as found in Ref.~\cite{MartinezTorres:2010ax}). Finally, we add that the opposite signs in the amplitudes shown in Fig.~\ref{compare} arise from the dynamics involved in the different subsystems. 

From the discussions made in this section, we can conclude that for the present system contributions from diagrams beyond those summed in the FCA series are negligible. Hence, we can conclude that the results obtained in the present work do not get significant corrections from the diagrams beyond FCA.

\section{Conclusions}
In this work we have investigated the formation of $N^*$ states as a consequence of the dynamics involved in the $ND\bar D^*-N\bar D D^*$ system. To do this, we solve the Faddeev equations treating the open charm mesons as a cluster. We find that the generation of $\Lambda_c(2595)$ in the $DN-D^*N$ system together with the clustering of $D$ ($D^*$) and $\bar D^*$ ($\bar D$) as $X(3872)$ or $Z_c(3900)$ produces enough attraction to form isospin $1/2$, $3/2$ states with masses in the energy region $4400-4600$ MeV and positive parity as summarized in Table~\ref{sum}, where the uncertainties are related to different models considered when determining the two-body interactions. The certainty of the results on isospin 3/2 states depends on the strength of the isovector $D^(*)N$ interactions, which are not well known yet.
\begin{table}[h!]
\caption{Summary of the isospin $1/2$ and $3/2$ states found in the present work.}\label{sum}
\begin{tabular}{cccc}
\hline
Isospin&Spin-parity&Mass (MeV)& Width (MeV)\\
\hline
$1/2$&$1/2^+$ & $4404-4410$ &2\\
$1/2$&$1/2^+$ & $4556-4560$ &$4-20$\\
$1/2$&$3/2^+$ & $4467-4513$ &$\sim 3-6$\\
$1/2$&$3/2^+$ & $4558-4565$ &$\sim 5-14$\\
$3/2$&$1/2^+$, $3/2^+$ & $4359$&$1.5$\\
$3/2$&$1/2^+$& $4512$& $4$\\
$3/2$&$3/2^+$& $4514$& $1$\\
\hline
\end{tabular}
\end{table}

In this way, we can conclude that $N^*$ and $\Delta^*$ states with hidden charm and positive parity arise from three-hadron
dynamics. We have discussed that data from $\Lambda_b$ decays are available on final states which can confirm the existence of such positive-parity states.

\section{Acknowledgements}
 This work is supported by the Funda\c c\~ao de Amparo \`a Pesquisa do Estado de S\~ao Paulo (FAPESP), processos n${}^\circ$ 2019/17149-3, 2019/16924-3 and 2020/00676-8, and by the Conselho Nacional de Desenvolvimento Cient\'ifico e Tecnol\'ogico (CNPq), grant  n${}^\circ$ 305526/2019-7 and 303945/2019-2. 
\clearpage
\bibliographystyle{unsrt}
\bibliography{refs}

\end{document}